\documentclass[preprint,aps]{revtex4}
\usepackage{epsf}
\usepackage{graphicx}% Include figure files
\usepackage{dcolumn}% Align table columns on decimal point
\usepackage{bm}% bold math
\everymath{\displaystyle}
\begin{document}

\title{ Quantum-Mechanical Description of the Electromagnetic Interaction
of Relativistic Particles with Electric and Magnetic Dipole Moments}

\author{
A. J. Silenko} \affiliation{Institute of Nuclear Problems,
Belarusian State University, 220080 Minsk, Belarus}
\email{silenko@inp.minsk.by}

\begin {abstract} The Hamiltonian of relativistic particles with electric and magnetic dipole moments that interact with
an electromagnetic field is determined in the Foldy-Wouthuysen
representation. Transition to the semiclassical approximation is
carried out. The quantum-mechanical and semiclassical equations of
spin motion are derived.
\end{abstract}
\maketitle
 \newpage

Nowadays the electric dipole moments of particles, nuclei, and atoms are intensively investigated. The components
comprising derivatives of the electric field strength cannot be neglected for atoms and nuclei, because they determine, in
particular, the contact (Darwin) interaction. Relativistic effects are also important for particles, nuclei, and heavy atoms.

A very convenient method of description of the relativistic particle interaction with an external field and of
transition to the semiclassical description is the Foldy-Wouthuysen transformation \cite{FW}.
In the Foldy-Wouthuysen
representation, the Hamiltonian and all operators have a block-diagonal structure (diagonal for two spinors). Relationships
between the operators are equivalent to those between the corresponding classical parameters. The operators in the Foldy-
Wouthuysen representation are the same as in nonrelativistic quantum theory. Only the Foldy-Wouthuysen representation
has these properties, which considerably simplify the transition to the semiclassical description. The Foldy-Wouthuysen
representation provides the best opportunity for the transition to the classical limit of relativistic quantum mechanics
\cite{FW,CMcK,JMP}.

The relativistic system of units with $\hbar=c=1$ is used in the
present work.

It is very important that in this representation, the operators of
coordinates $\bm r$, momentum
 $\bm p=-i\nabla$, and polarization
$$\bm\Pi=\left(\begin{array}{cc} \bm{\sigma}  &  0 \\ 0 &
-\bm{\sigma}\end{array}\right), $$ где $\bm{\sigma}$
have very simple forms. Here $\bm{\sigma}$ is the Pauli matrix. In other representations, these operators are described by much more
cumbersome formulas \cite{FW,JMP}. This makes the Foldy-Wouthuysen representation extremely convenient for obtaining
equations of particle and spin motion. In particular, the operator equation of spin motion is given by the formula
\begin{equation} \frac{d\bm\Pi}{dt}=i[{\cal H},\bm\Pi],
\label{eq1} \end{equation} where ${\cal H}$ is the Hamiltonian, and $
[\dots, \dots]$ designates the commutator. To derive the semiclassical equation of spin motion,
averaging over wave functions [3] must be performed.

In [3] the Hamiltonian of relativistic particles with a spin of 1/2 interacting with the electromagnetic field was
found in the Foldy-Wouthuysen representation. Calculations were carried out to within derivatives of electric and magnetic
external field strengths (the fields were generally nonstationary) with allowance for the anomalous particle magnetic
moments.

In the present work, the Foldy-Wouthuysen transformation is used for relativistic particles with anomalous
magnetic and electric dipole moments interacting with an electromagnetic field. The transformation method was described
in detail in [3].

The electric dipole moment can be considered by inclusion of terms characterising the electric dipole moment into
the Dirac-Pauli Hamiltonian describing the interaction of particles having anomalous magnetic moments with the
electromagnetic field. The Dirac-Pauli equation has the form
\begin{equation} \biggl[\gamma^\mu\pi_\mu-m+\frac{\mu'}{2}\sigma^{\mu\nu}F_{\mu\nu}\biggr]
\Psi=0, ~~~~~ \pi_\mu=p_\mu-eA_\mu,\label{eq2} \end{equation}
where    $\gamma^\mu$ are the Dirac matrices, $F_{\mu\nu}$
is the electromagnetic field tensor;
$p^\mu$ and $A^\mu=(\Phi,\bm A)$ are the four-dimensional
particle momentum and external field potential, respectively;
$\sigma^{\mu\nu}=i(\gamma^\mu\gamma^\nu-\gamma^\nu\gamma^\mu)/2$, $m$ is the mass of the particle at rest;
and  $\mu'$ is the anomalous particle magnetic moment.
The corresponding Hamiltonian in the Foldy-Wouthuysen representation disregarding the electric dipole moment is
characterized by the expression \cite{JMP}:
\begin{equation}\begin{array}{c} {\cal H}=\beta\epsilon'+e\Phi+\frac
   14\left\{\left(\frac{\mu_0m}{\epsilon'
   +m}+\mu'\right)\frac{1}{\epsilon'},\biggl(\bm\Sigma\!\cdot\![\bm\pi\!
\times\!\bm E]-\bm\Sigma\!\cdot\![\bm E\!\times\!\bm\pi]-\nabla\!
\cdot\!\bm E\biggr)\right\}_+\\ +\frac{\mu_0m}{16}
\left\{\frac{2\epsilon'^2+2\epsilon' m+m^2}{\epsilon'^4( \epsilon'
+m)^2},\bm\pi\!\cdot\!\nabla(\bm\pi\!\cdot\!\bm E+\bm
E\!\cdot\!\bm\pi)\right\}_+-\frac
12\left\{\left(\frac{\mu_0m}{\epsilon'}
+\mu'\right), \bm\Pi\!\cdot\!\bm H\right\}_+\\
+\frac{\mu'}{4}\left\{\frac{1}{\epsilon'(\epsilon'+m)},
\biggl[(\bm{H}\!\cdot\!\bm\pi)(\bm{\Pi}\!\cdot\!\bm\pi)+ (\bm{\Pi}
\!\cdot\!\bm\pi)(\bm\pi\!\cdot\!\bm{H})+2\pi(\bm\pi\!\cdot\!\bm j+
\bm j\!\cdot\! \bm\pi)\biggr]\right\}_+,
\end{array} \label{eq3} \end{equation}
where $\{\dots,\dots\}_+$ designates the anticommutator, $\mu_0=e/(2m)$
is the Dirac magnetic moment, and
\begin{equation}
\epsilon'=\sqrt{m^2+\bm{\pi}^2}.\label{eq4} \end{equation}

The anomalous magnetic and electric dipole moments are closely related, because they determine the real and
imaginary parts of the same physical quantity
\cite{FMS,GTh}. The contributions of the anomalous magnetic and electric dipole
moments to the Lagrangian are the following \cite{FMS}:
\begin{equation} {\cal L}_{AMM}=\frac{\mu'}{2}\sigma^{\mu\nu}F_{\mu\nu},
~~~~~ {\cal L}_{EDM}=-i\frac{d}{2}\sigma^{\mu\nu}\gamma^5F_{\mu\nu}, ~~~~~
 \gamma^5=\left(\begin{array}{cc} 0  &  -1 \\ -1 & 0 \end{array}\right),
\label{eq5} \end{equation} where $d$ is the electric dipole moment
of the particle, and 0 and $-1$ designate the corresponding
$2\times2$ matrices.

A consideration of the electric dipole moment of the particle consists in the inclusion of the term proportional to
$d$
into the Dirac-Pauli equation. As a result, this equation assumes the form
\begin{equation} \biggl[\gamma^\mu\pi_\mu-m+\frac{\mu'}{2}\sigma^{\mu\nu}F_{\mu\nu}
-i\frac{d}{2}\sigma^{\mu\nu}\gamma^5 F_{\mu\nu}\biggr]
\Psi=0.\label{eq6} \end{equation}

The Hamiltonian in the Dirac representation is described by the expression
\begin{equation}
{\cal H}_D=\beta m+\bm\alpha\cdot\bm\pi+e\Phi+\mu'(-\bm\Pi\cdot\bm H+
i\bm\gamma\cdot\bm E)-id(-\bm\Pi\cdot\bm H+
i\bm\gamma\cdot\bm E)\gamma^5,\label{eq7} \end{equation}
where $\bm\pi=-i\nabla-e\bm A$ and  $\bm E$ and
$\bm H$ are the electric and magnetic field strengths. Hereinafter, we use the following standard
designations:
$$\begin{array}{c}\bm{\gamma}=\left(\begin{array}{cc} 0  &  \bm{\sigma} \\ -\bm{\sigma} & 0
\end{array}\right), ~~~ {\beta}\equiv\gamma^0=\left(\begin{array}{cc} 1  &  0
\\ 0 & -1 \end{array}\right), ~~~\bm{\alpha}=\beta\bm\gamma=
\left(\begin{array}{cc} 0  &  \bm{\sigma} \\ \bm{\sigma} & 0
\end{array}\right), \\    \bm{\Sigma}
=\left(\begin{array}{cc} \bm{\sigma}  &  0 \\ 0 &
\bm{\sigma}\end{array}\right),   ~~~\bm{\Pi}=\beta\bm\Sigma
=\left(\begin{array}{cc} \bm{\sigma}  &  0 \\ 0 &
-\bm{\sigma}\end{array}\right).  \end{array}  $$

The Hamiltonian given by Eq. (7) is reduced to the form
\begin{equation}
{\cal H}_D=\beta m+\bm\alpha\cdot\bm\pi+e\Phi+\mu'(-\bm\Pi\cdot\bm H+
i\bm\gamma\cdot\bm E)-d(\bm\Pi\cdot\bm E+
i\bm\gamma\cdot\bm H).
\label{eq8} \end{equation}

Formula (8) demonstrates that terms describing contributions of anomalous magnetic and electric dipole moments
to the Hamiltonian are transformed into each other using the substitutions
\begin{equation} \bm
H\rightarrow \bm E,~ \bm E\rightarrow-\bm H,~
\mu'\rightarrow d.\label{eq9} \end{equation}
The same relationships between these terms take place in the classical description.

We note here that to correctly include the electric dipole moment into the Dirac-Pauli equation, we can obviate the
need for Eqs. (5) and (6). A more natural form is used in classical electrodynamics [6]. In this case, the interaction of the
electric dipole moment with the electromagnetic field is described by the tensor
 $G^{\mu\nu}=(-\bm H,-\bm E)$ dual with respect to the
electromagnetic field one $F^{\mu\nu}=(-\bm E,\bm H)$.
When the tensor $G^{\mu\nu}$ is used to describe the electric dipole moment, the generalized
Dirac-Pauli equation assumes the form
\begin{equation} \biggl[\gamma^\mu\pi_\mu-m+\frac{\mu'}{2}\sigma^{\mu\nu}
F_{\mu\nu}-\frac{d}{2}\sigma^{\mu\nu}G_{\mu\nu}\biggr]
\Psi=0,\label{eq10} \end{equation}
and the Lagrangian ${\cal L}_{EDM}$
\begin{equation}
{\cal L}_{EDM}=-\frac{d}{2}\sigma^{\mu\nu}G_{\mu\nu}.
\label{eq11} \end{equation}

The Hamiltonian of particles with anomalous magnetic and electric dipole moments, calculated by the method
suggested in [3], has the form
\begin{equation}\begin{array}{c} {\cal H}=\beta\epsilon'+e\Phi+\frac
   14\left\{\left(\frac{\mu_0m}{\epsilon'
   +m}+\mu'\right)\frac{1}{\epsilon'},\biggl(\bm\Sigma\!\cdot\![\bm\pi\!
\times\!\bm E]-\bm\Sigma\!\cdot\![\bm E\!\times\!\bm\pi]-\nabla\!
\cdot\!\bm E\biggr)\right\}_+\\ +\frac{\mu_0m}{16}
\left\{\frac{2\epsilon'^2+2\epsilon' m+m^2}{\epsilon'^4( \epsilon'
+m)^2},\bm\pi\!\cdot\!\nabla(\bm\pi\!\cdot\!\bm E+\bm
E\!\cdot\!\bm\pi)\right\}_+-\frac
12\left\{\left(\frac{\mu_0m}{\epsilon'}
+\mu'\right), \bm\Pi\!\cdot\!\bm H\right\}_+\\
+\frac{\mu'}{4}\left\{\frac{1}{\epsilon'(\epsilon'+m)},
\biggl[(\bm{H}\!\cdot\!\bm\pi)(\bm{\Pi}\!\cdot\!\bm\pi)+ (\bm{\Pi}
\!\cdot\!\bm\pi)(\bm\pi\!\cdot\!\bm{H})+2\pi(\bm\pi\!\cdot\!\bm j+
\bm j\!\cdot\! \bm\pi)\biggr]\right\}_+
-d\bm\Pi\!\cdot\!\bm E\\ +\frac{d}{4}\left\{\frac{1}{\epsilon'(\epsilon'+m)},
\biggl[(\bm{E}\!\cdot\!\bm\pi)(\bm{\Pi}\!\cdot\!\bm\pi)+ (\bm{\Pi}
\!\cdot\!\bm\pi)(\bm\pi\!\cdot\!\bm{E})\biggr]\right\}_+
-\frac d4\left\{\frac{1}{\epsilon'},\biggl(\bm\Sigma\!\cdot\![\bm\pi\!
\times\!\bm H]-\bm\Sigma\!\cdot\![\bm H\!\times\!\bm\pi]\biggr)\right\}_+.
\end{array} \label{eq12} \end{equation}

A comparison between Eqs. (3) and (12) demonstrates that substitution (9) is allowable for Hamiltonian (3).
Despite analogous descriptions of the anomalous magnetic and electric dipole moments, the interaction of these two
moments with the electromagnetic field differs in principle. In the expression for the Hamiltonian in the Foldy-Wouthuysen
representation, there are no terms proportional to the electric dipole moment and involving the first derivatives of the field
strengths. The corresponding terms proportional to the anomalous magnetic moment characterize the contact interaction
with external charges and currents. This result, caused by the absence of magnetic charges and currents, is very important,
because it simplifies the estimation of the electric dipole moment contribution to relativistic expression (12) for the
Hamiltonian.

An analysis of Eq. (12) demonstrates that in the examined approximation, the last but one term gives the main
relativistic correction to the operator describing the interaction of the electric dipole moment with the external field for
atoms and nuclei:
\begin{equation} \Delta{\cal H}=
\frac{d}{4}\left\{\frac{1}{\epsilon'(\epsilon'+m)},
\biggl[(\bm{E}\!\cdot\!\bm\pi)(\bm{\Pi}\!\cdot\!\bm\pi)+ (\bm{\Pi}
\!\cdot\!\bm\pi)(\bm\pi\!\cdot\!\bm{E})\biggr]\right\}_+.
\label{eq13} \end{equation}

Let us emphasize that the relativistic corrections for nuclei and heavy atoms are not small and can reach several
tens of percent [7]. Their calculations in the Foldy-Wouthuysen representation (rather than in the Dirac representation
commonly used) can be more convenient.

The particle spin motion in the electromagnetic field is also very important. The operator equation of spin motion
for relativistic particles with anomalous magnetic and electric dipole moments, derived with the help of Eqs. (1) and (12),
has the form
\begin{equation}
\begin{array}{c}
\frac{d\bm{\Pi}}{dt} =\left\{\left(\frac{\mu_0m} {\epsilon
'+m}+\mu'\right)\frac{1}{\epsilon '},\left[\bm{\Pi}\times[\bm
E\times\bm \pi]\right]\right\}_++ \left\{\left(
\frac{\mu_0m}{\epsilon '}+\mu'\right),[\bm\Sigma\times\bm
H]\right\}_+ \\ -\frac{\mu'}{2}\left\{\frac{1}{\epsilon '(\epsilon
'+m)},\biggl( [\bm\Sigma\times\bm \pi](\bm \pi\cdot\bm H)+(\bm
H\cdot\bm \pi) [\bm\Sigma\times\bm
\pi]\biggr)\right\}_++2d[\bm\Sigma\times\bm
E]\\-\frac{d}{2}\left\{\frac{1}{\epsilon '(\epsilon '+m)},\biggl(
[\bm\Sigma\times\bm \pi](\bm \pi\cdot\bm E)+(\bm E\cdot\bm \pi)
[\bm\Sigma\times\bm \pi]\biggr)\right\}_+
-d\left\{\frac{1}{\epsilon '},\left[\bm{\Pi}\times[\bm H\times\bm
\pi]\right]\right\}_+.
\end{array} \label{eq14} \end{equation}

Transition to the semiclassical approximation was described in [3]. Introducing the factor
$\eta=4dm/e$, we obtain the
semiclassical equation of spin motion:
\begin{equation}\begin{array}{c} \frac{d\bm \xi}{dt}=(\bm\Omega_{BMT}+\bm\Omega_{EDM})\times\bm
\xi,\\ \bm\Omega_{BMT}=-\frac{e}{2m}
\left\{\left(g-2+\frac{2}{\gamma}\right)\bm H-
\frac{(g-2)\gamma}{\gamma+1}\bm\beta(\bm \beta\cdot \bm H)
%    \right. \\ \left.
-\left(g-2+\frac{2}{\gamma+1}\right)[\bm\beta\times\bm E]\right\},
\\
\bm\Omega_{EDM}=-\frac{e\eta}{2m}\left(\bm
E-\frac{\gamma}{\gamma+1}\bm\beta(\bm\beta\cdot\bm
E)+\bm\beta\times\bm
H\right),\end{array}\label{eq15}\end{equation} where
$\bm\Omega_{BMT}$ is determined by the Thomas-Bargmann-Michel-Telegdi equation [8]. Equation (15) coincides with the
corresponding classical equation [6].

The influence of the electric dipole moment on the particle motion is negligibly small; however, it influences
significantly the spin motion. The character of spin motion caused by the interaction of the electric and magnetic dipole
moments with the external field differs radically. This circumstance allows experiments on measuring the electric dipole
moment in rings with memory to be carried out [9]. The spin rotation about the momentum vector in the horizontal plane
can be compensated with the help of application of a radial electric field having the strength
$$\bm E=\frac{a\gamma^2}{1-a\beta^2\gamma^2}[\bm\beta\times\bm H].$$
In this case, the contribution of the electric dipole moment to
the spin motion is characterized by the angular velocity
\begin{equation}\begin{array}{c}
\bm\Omega_{EDM}=-\frac{e\eta}{2m}\cdot\frac{1+a}{1-a\beta^2\gamma^2}[\bm\beta\times\bm
H].\end{array}\label{eq16}\end{equation} The difference in signs
with [9] is explained by the fact that in the present work, the
angular velocity vector was defined with the opposite sign.

Thus, the Foldy-Wouthuysen transformation has been successively
used to determine the Hamiltonian and the equation of spin motion
for relativistic particles with anomalous magnetic and electric
dipole moments. Relativistic corrections to the interaction
operator can be used to calculate the electric dipole moments of
nuclei and heavy atoms.

\end{document}